\documentclass[preprintnumbers]{revtex4}

\usepackage{graphicx}
\usepackage{epsf}
\usepackage{amsmath,amssymb}
\newcommand{\bvec}{\boldsymbol}

\begin{document}
\title{Isoscalar monopole and dipole excitations of cluster states and giant resonances in $^{12}$C}
\author{Yoshiko Kanada-En'yo}
\affiliation{Department of Physics, Kyoto University, Kyoto 606-8502, Japan}
\begin{abstract}
The isoscalar monopole(ISM) and dipole(ISD) excitations in $^{12}$C are theoretically investigated
with the shifted antisymmetrized molecular dynamics(AMD) plus $3\alpha$-cluster generator coordinate method(GCM). 
The small amplitude vibration modes are described by coherent one-particle one-hole excitations 
expressed by small shift of single-nucleon 
Gaussian wave functions within the AMD framework, whereas the large amplitude cluster modes are incorporated 
by superposing $3\alpha$-cluster wave functions in the GCM. The coupling of the excitations
in the intrinsic frame with the rotation and parity transformation is taken into account 
microscopically by the angular-momentum and parity projections.
The present calculation describes the ISM and ISD excitations 
in a wide energy region covering cluster modes in the low-energy region 
and the giant resonances in the high-energy region, though the quantitative description of the
high-energy part is not satisfactory. 
The low-energy ISM and ISD strengths of the cluster modes are 
enhanced by the radial motion of $\alpha$ clusters, and they split into
a couple of states because of the angular motion of $\alpha$ clusters.
The low-energy ISM strengths
exhaust 26\% of the EWSR, which is consistent with the experimental data
for the $^{12}$C($0^+_2$;7.65 MeV) and $^{12}$C($0^+_3$;10.3 MeV) 
measured by $(e,e')$, $(\alpha,\alpha')$, and ($^6$Li,$^6$Li$'$) 
scatterings.
In the calculated low-energy ISD strengths, two  $1^-$ states (the $1^-_1$ and $1^-_2$) 
with the significant strengths are obtained in $E=10-15$ MeV.
It is indicated that the ISD excitations
can be a good probe to experimentally search for new cluster states such as the
$^{12}$C($1^-_2$) obtained in the present calculation. 
\end{abstract}
\maketitle
\section{Introduction}
Isoscalar monopole(ISM) and dipole(ISD) strength 
distributions measured by hadron inelastic scattering experiments
have been providing useful informations for nuclear properties concerning nuclear excitations
corresponding to the compressional vibration modes
as well as the nuclear matter incompressibility.  
Isoscalar giant monopole resonances(ISGMRs), which were
established in medium and heavy nuclei \cite{Youngblood:1977zz,Morsch:1979zz,Rozsa:1980zz,Youngblood:1981zz}, 
have been studied extensively between $^{12}$C and $^{206}$Pb \cite{Youngblood:1998zz,Youngblood:1999zz,Lui:2001xh,Youngblood:2001mq,John:2003ke,Chen:2009zzp}.
The ISGMRs in medium and heavy nuclei exhaust almost 100\% of the isoscalar monopole(ISM) 
energy-weighted sum rule(EWSR), and the excitation energies 
have been used to extract the compressibility of the nuclear matter with microscopic calculations using, 
for example, non-relativistic and relativistic mean-field approaches and those plus random-phase approximation(RPA)
(see Refs.~\cite{Blaizot:1976cw,Blaizot:1980tw,Vretenar:2003qm,Colo:2004mj,Paar:2007bk} and references therein).
High precision studies using hadron inelastic scatterings have revealed that
the ISM strength distributions in nuclei lighter than $^{40}$Ca 
are strongly fragmented
\cite{Youngblood:1998zz,Youngblood:1999zz,Lui:2001xh,John:2003ke,Chen:2009zzp}.
In the ISM strengths in $^{16}$O, the significant percentage of the EWSR has been found in a low-energy region. 
In the theoretical study of $^{16}$O with a $4\alpha$ calculation by
Yamada {\it et al.}, it was pointed out that two different types of ISM excitations exist in $^{16}$O
\cite{Yamada:2011ri}:
the low-energy ISM strengths of cluster excitations in $E\lesssim 16$ MeV 
are separated from 
the high-energy ISM strengths of the ISGMR in $E> 16$ MeV, which correspond to 
the collective vibration mode described by coherent one-particle and one-hole (1p-1h) excitations in a mean field.
The separation of the low-energy ISM strengths from the ISGMR was also found
in $^{12}$C \cite{Youngblood:1998zz,John:2003ke}. The ISM strength distributions in $^{12}$C 
observed by $(\alpha,\alpha')$ and 
$(^6\textrm{Li}, ^6\textrm{Li}')$ scatterings \cite{Youngblood:1998zz,John:2003ke,Eyrich:1987zz}
show that the low-energy ISM strengths in $E\lesssim 12$ MeV
exhaust the significant percentage of the EWSR comparable to the high-energy strengths 
in $E> 12$ MeV of the ISGMR. In $^{12}$C, a couple of excited $0^+$ states have been experimentally 
observed near the $3\alpha$ threshold energy (see, for instance, Refs.~\cite{John:2003ke,Itoh:2011zz,Freer:2014qoa} and references therein). 
Theoretically, these states are considered to be $3\alpha$-cluster states \cite{kamimura-RGM1,kamimura-RGM2,uegaki1,uegaki3,descouvemont87,KanadaEn'yo:1998rf,Tohsaki01,Funaki:2003af,Neff-c12,
Fedotov:2004nz,kurokawa04,Kurokawa:2005ax,Filikhin:2005nc,KanadaEn'yo:2006ze,
Arai:2006bt,Funaki:2006gt,Chernykh:2007zz,ohtsubo13,Ishikawa:2014mza}. 
The remarkable E0 
strength for the $0^+_2$(7.65 MeV) state was observed by electron scattering \cite{strehl70,Chernykh:2010zu}, 
and the inelastic form factor for the $0^+_2$ is described well by $3\alpha$-cluster models \cite{kamimura-RGM2,uegaki3,Filikhin:2005nc,Chernykh:2007zz}.
In general, the ISM strengths can be strongly concentrated on cluster states in the 
low-energy region
because the ISM operator directly excites the cluster motion with $L=0$ as well as the compressional monopole vibration mode \cite{Yamada:2011ri}. 
Therefore, the low-energy ISM strengths separating from the ISMGR is a good probe 
to study cluster states in light nuclei
as discussed for $^{12}$C, $^{16}$O, $^{11}$B, and $^{24}$Mg 
\cite{Yamada:2011ri, Funaki:2006gt,Kawabata:2005ta,KanadaEn'yo:2006bd,Wakasa:2007zza,
Ichikawa:2011ji,kawabata13,Chiba:2015zxa}. 

In analogy to the ISM excitations, the ISD excitations can be another probe to search for 
cluster states because the ISD operator is able to excite the cluster motion with $L=1$
as well as the compressional dipole vibration mode.
The isoscalar giant dipole resonances (ISGDRs) observed by $(\alpha,\alpha')$ scatterings
are found generally in the higher energy region than the ISGMR 
(see Refs.~\cite{Paar:2007bk,morsch80,Harakeh:1981zz,Colo:2003zm} and references therein). 
Compared with the ISM strength distributions, the observed ISD strengths in medium and 
heavy nuclei are strongly fragmented 
and concentrate typically in two bump structures \cite{Clark:2001rb,Lui:2006zk}: one in the low-energy region 
and the other in the relatively higher energy region, which are considered to correspond to
the toroidal dipole mode and the compressional dipole vibration mode, respectively 
\cite{Paar:2007bk,Colo:2003zm,giai81,Colo:2000br,Vretenar:2001te,Papakonstantinou:2010ja}. 
In $^{16}$O and $^{40}$Ca, the low-energy ISD strengths 
concentrate on the $1^-$ states at $E=7.12$ MeV and $E=6.95$ MeV, respectively, 
which exhaust approximately 4\% of the EWSR \cite{Harakeh:1981zz,poelhekken92}. 
Also in $^{12}$C, the significant low-energy ISD strengths have been observed in $E\le 20$ MeV
below the high-energy strengths of the ISGDR \cite{John:2003ke}.
The separation of the low-energy ISD strengths from the high-energy strengths of the ISGDR for the 
compressional dipole vibration mode suggests
two different types of ISD modes. In the case of light nuclei, the low-energy ISD strengths
could correspond to the cluster excitations.

The aim of this paper is to investigate ISM and ISD excitations in $^{12}$C from low-energy to
high-energy regions and to clarify the separation of the cluster excitations from the 
 ISGMR and ISGDR for the collective vibration modes. 
In the history of theoretical studies of the giant resonances, 
RPA calculations in mean-field approaches have been widely applied.
Although the RPA calculations are successful for a variety of collective 
excitations in heavy mass nuclei, the RPA is not useful to describe well-developed 
cluster states and fails to reproduce  
the low-energy ISM strengths in $^{16}$O \cite{Lui:2001xh,Ma:1997zzb}
because it is a small amplitude approach and unable to describe the large amplitude cluster motions.
Moreover, in most of current mean-field approaches, 
calculations are based a parity-symmetric mean field in a strong coupling picture
without the angular-momentum and parity projections, and the 
coupling of single-particle excitations in the mean-field with the rotation and parity transformation is 
not taken into account microscopically.
The ISM transitions to the $0^+$ states having $3\alpha$-cluster structures near and 
above the $3\alpha$ threshold have been theoretically investigated by microscopic 
and semi-microscopic $3\alpha$-cluster models 
\cite{kamimura-RGM2,uegaki3,Filikhin:2005nc,Funaki:2006gt} as well as other structure models such as the
antisymmetrized molecular dynamics (AMD) \cite{KanadaEn'yo:2006ze} and fermionic molecular dynamics (FMD) \cite{Chernykh:2007zz} and also the 
{\it ab initio} calculation \cite{Lovato:2013cua}.
However, these calculations have not yet been applied to investigate 
the relatively high-energy ISGMR and ISGDR strengths for the collective vibration modes 
contributed by coherent 1p-1h excitations.

To take into account  the large amplitude cluster modes and
the coherent 1p-1h excitations as well as the angular-momentum and parity projections, I have recently 
developed the AMD method \cite{KanadaEnyo:1995tb,KanadaEnyo:1995ir,AMDsupp,KanadaEn'yo:2012bj} in Refs.~\cite{Kanada-En'yo:2013dma,Kanada-En'yo:2015ttw}. 
The applications of the AMD to collective vibration modes go back to 
the works on electric dipole (E1) and ISM excitations \cite{KanadaEn'yo:2005wd,Furuta:2010ad}
with the time-dependent AMD,  which was originally developed 
for study of heavy-ion reactions \cite{Ono:1991uz,Ono:1992uy}.
However, in the time-dependent AMD approach,
the angular-momentum and parity projections are not performed. 
Instead of the time-dependent AMD, 
we superpose the angular-momentum and parity projected wave functions of 
various configurations including the 1p-1h and cluster excitations. 
I first perform the variation after the angular-momentum and parity projections in the AMD 
framework (AMD+VAP) to obtain the ground state wave function of $^{12}$C as done in Refs.~\cite{KanadaEn'yo:1998rf,KanadaEn'yo:2006ze}. 
Then I describe small amplitude motions by taking into 
account 1p-1h excitations on the obtained ground 
state wave function with the shifted AMD method \cite{Kanada-En'yo:2013dma,Kanada-En'yo:2015ttw}.
To incorporate the large amplitude cluster motions, I combine the 
the generator coordinate method (GCM) with the shifted AMD
by superposing $3\alpha$-cluster wave functions. 
The angular-momentum and parity projections are performed in the present framework.
Applying the present method, I investigate the ISM and ISD excitations in $^{12}$C.

This paper is organized as follows. 
The present method is formulated in section \ref{sec:formulation}, 
and section \ref{sec:results} discusses the ground state structure and the ISM and ISD 
excitations in $^{12}$C. 
The paper concludes with a summary in section \ref{sec:summary}.

\section{Formulations of shifted AMD and $3\alpha$-cluster GCM for ISM and ISD excitations}
\label{sec:formulation}
In order to calculate the ISM and ISD excitations in $^{12}$C, 
I apply the shifted AMD based on the AMD+VAP and combine it with the $3\alpha$-cluster GCM. 
In this section, I explain the formulations of the AMD+VAP,  the shifted AMD, and the $3\alpha$-cluster GCM, 
and also describe the definitions of the ISM and ISD transitions.
For the details of the AMD method, the reader is referred to Refs.~\cite{AMDsupp,KanadaEn'yo:2012bj} 
and references therein. 

 \subsection{AMD wave function}
An AMD wave function is given by a Slater determinant,
\begin{equation}
 \Phi_{\rm AMD}({\bvec{Z}}) = \frac{1}{\sqrt{A!}} {\cal{A}} \{
  \varphi_1,\varphi_2,...,\varphi_A \},\label{eq:slater}
\end{equation}
where  ${\cal{A}}$ is the antisymmetrizer. The $i$th single-particle wave function 
$\varphi_i$ is written by a product of 
spatial, spin, and isospin
wave functions as
\begin{eqnarray}
 \varphi_i&=& \phi_{{\bvec{X}}_i}\chi_i\tau_i,\\
 \phi_{{\bvec{X}}_i}({\bvec{r}}_j) & = &  \left(\frac{2\nu}{\pi}\right)^{4/3}
\exp\bigl\{-\nu({\bvec{r}}_j-\bvec{X}_i)^2\bigr\},
\label{eq:spatial}\\
 \chi_i &=& (\frac{1}{2}+\xi_i)\chi_{\uparrow}
 + (\frac{1}{2}-\xi_i)\chi_{\downarrow}.
\end{eqnarray}
$\phi_{{\bvec{X}}_i}$ and $\chi_i$ are the spatial and spin functions, respectively, and 
$\tau_i$ is the isospin
function fixed to be up (proton) or down (neutron). 
The width parameter $\nu$ is fixed to be the optimized value. 
To separate the center of mass motion from the total wave function $\Phi_{\rm AMD}({\bvec{Z}})$,
the following condition should be satisfied,  
\begin{equation}\label{eq:cm}
\frac{1}{A}\sum_{i=1,\ldots,A} \bvec{X}_i=0.
\end{equation}
In the present calculation, I keep this condition and exactly remove the contribution of
the center of mass motion.
Accordingly, an AMD wave function
is expressed by a set of variational parameters, ${\bvec{Z}}\equiv 
\{{\bvec{X}}_1,\ldots, {\bvec{X}}_A,\xi_1,\ldots,\xi_A \}$,
which specify centroids of single-nucleon Gaussian wave packets and spin orientations 
for all nucleons. It should be commented that the AMD wave function is similar to the wave 
function used in FMD calculations \cite{Feldmeier:1994he,Neff:2002nu}.

In the AMD framework, existence of clusters is not assumed {\it a priori} 
because Gaussian centroids, ${\bvec{X}}_1,\ldots,{\bvec{X}}_A$, of
all single-nucleon wave packets are independently 
treated as variational parameters. Nevertheless, 
a multi-center cluster wave function can be described by the 
AMD wave function with the corresponding configuration of Gaussian centroids. 
It should be also commented that, when all $|\bvec{X}_i|$ goes to the small limit, 
the AMD wave function 
is equivalent to a shell-model configuration because of the mathematical consequence of
the antisymmetrized Gaussian wave packets.


\subsection{AMD+VAP}
To obtain the ground state wave function of an $A$-nucleon system,
the AMD+VAP method is applied. 
In the AMD+VAP method, the parameters ${\bvec{Z}}= 
\{{\bvec{X}}_1,{\bvec{X}}_2,\ldots, {\bvec{X}}_A,\xi_1,\xi_2,\ldots,\xi_A \}$
in the AMD wave function
are determined by the energy variation 
after the angular-momentum and parity projections (VAP).
It means that 
${\bvec{X}}_i$ and $\xi_{i}$ for the lowest $J^\pi$ state 
are optimized so as to minimize the energy expectation value of the Hamiltonian
for the $J^\pi$-projected AMD wave function; 
\begin{eqnarray}
&& \frac{\delta}{\delta{\bvec{X}}_i}
\frac{\langle \Phi|H|\Phi\rangle}{\langle \Phi|\Phi\rangle}=0,\\
&& \frac{\delta}{\delta\xi_i}
\frac{\langle \Phi|H|\Phi\rangle}{\langle \Phi|\Phi\rangle}=0,\\
&&\Phi= P^{J\pi}_{MK}\Phi_{\rm AMD}({\bvec{Z}}),
\end{eqnarray}
where $P^{J\pi}_{MK}$ is the angular-momentum and parity projection operator. 
After the VAP calculation with $J^\pi=0^+$, 
the optimized parameters $\bvec{Z}^0=\{\bvec{X}^0_1,\ldots,\xi^0_{1},\ldots\}$
for the ground state of $^{12}$C are obtained. 

\subsection{Shifted AMD}
To incorporate 1p-1h excitations, 
I consider small variation $\delta \varphi_i$ of the 
single-particle wave function $\varphi_i$ in the 
ground state wave function $\Phi_{\rm AMD}({\bvec{Z}^0})$
by shifting the position of the Gaussian centroid,
${\bvec{X}}^0_i\rightarrow {\bvec{X}}^0_i+\bvec{\Delta}_\sigma$, where
$\bvec{\Delta}_\sigma$ is the small vector specified by the label $\sigma$ for the shift direction. 
In the present calculation, 
8 directions ($\sigma=1,\ldots,8$) are adopted to obtain the approximately converged result 
for the ISM and ISD strengths. I choose enough small shift, typically 
$\Delta=|\bvec{\Delta}_\sigma|=0.1\sim 0.2$ fm, so as 
to obtain $\Delta$-independent results.
Details of the adopted ${\bvec{\Delta}}_{\sigma=1,\ldots,8}$ are described 
in section \ref{sec:results}.
For the spin part, 
I consider the spin-nonflip single-particle state $\chi_i$
and the spin-flip state $\bar\chi_i$ ($\langle\bar\chi_i|\chi_i\rangle=0$);
\begin{equation}
\bar\chi_i = (\frac{1}{2}+\bar\xi_i)\chi_{\uparrow}
 + (\frac{1}{2}-\bar\xi_i)\chi_{\downarrow},
\end{equation} 
where $\bar\xi_i=-1/4\xi^*_i$.
For all single-particle wave functions, I consider spin-nonflip and spin-flip
states shifted to 8 directions $({\bvec{\Delta}}_{\sigma=1,\ldots,8})$ independently and prepare 
$16A$ AMD wave functions, 
$\Phi_{\rm AMD}({\bvec{Z}}_{\rm nonflip}^0(i,\sigma))$
and $\Phi_{\rm AMD}({\bvec{Z}}_{\rm flip}^0(i,\sigma))$ with the parameters,
\begin{eqnarray}
&&\bvec{Z}_{\rm nonflip}^0(i,\sigma)\equiv 
\{{\bvec{X}^0_1}',\cdots,{\bvec{X}^0_i}'+ {\bvec{\Delta}}_\sigma,\cdots,
{\bvec{X}^0_A}',\xi^0_1,\cdots,\xi^0_i,\cdots,\xi^0_A \},\\
&&\bvec{Z}_{\rm flip}^0(i,\sigma)\equiv 
\{{\bvec{X}^0_1}',\cdots,{\bvec{X}^0_i}'+{\bvec{\Delta}}_\sigma,\cdots,
{\bvec{X}^0_A}',\xi^0_1,\cdots,\bar\xi^0_i,\cdots,\xi^0_A \},
\end{eqnarray}
where ${{\bvec{X}}^{0}_j}'={\bvec{X}}^{0}_j-{\bvec{\Delta}}_\sigma/(A-1)$ to
 take into account the recoil effect so that 
the center of mass motion is separated exactly.
Those shifted AMD wave functions 
$\Phi_{\rm AMD}({\bf Z}_{\rm (non)flip}^0(i,\sigma))$ 
and the original wave function
$\Phi_{\rm AMD}({\bf Z}^0)$
are superposed to obtain the final wave functions for the ground and excited states, 
\begin{eqnarray}
\Psi^{\rm sAMD}_{^{12}\textrm{C}(J^\pi_k)}&=&\sum_K c_0(J^\pi_k; K) 
P^{J\pi}_{MK}\Phi_{\rm AMD}({\bf Z}^0)\nonumber\\
&&+\sum_{i=1,\ldots,A}\sum_{\sigma}\sum_K 
c_1(J^\pi_k; i,\sigma,K) P^{J\pi}_{MK}\Phi_{\rm AMD}({\bf Z}_{\rm nonflip}^0(i,\sigma))\nonumber\\
&&+
\sum_{i=1,\ldots,A}\sum_{\sigma}\sum_K 
c_2(J^\pi_k; i,\sigma,K) P^{J\pi}_{MK}\Phi_{\rm AMD}({\bf Z}_{\rm flip}^0(i,\sigma)), \label{eq:sAMD}
\end{eqnarray}
where the coefficients $c_0$, $c_1$, and $c_2$ are determined by diagonalization 
of the norm and Hamiltonian matrices.
I call this method ``the shifted AMD'' (sAMD). 

The model space of the sAMD contains
the 1p-1h excitations that
are written by the small shift of 
single-nucleon Gaussian wave functions
of the ground state wave function. In the intrinsic frame 
before the angular-momentum and parity projections, the ground state AMD wave function is expressed by 
a Slater determinant, and therefore,  
the sAMD method corresponds to the RPA
in the restricted model space of the linear combination of the shifted Gaussian wave functions. 
However, since the $J^\pi$-projected wave functions are 
superposed in the sAMD, the coupling of the
1p-1h excitations with the rotation and parity transformation
is taken into account microscopically. Moreover, the original wave function for the 
ground state is the one obtained by the variation after the angular-momentum 
and parity projections as mentioned previously. 
It means that the sAMD may contain, in principle,
higher correlations beyond the RPA.

\subsection{$3\alpha$-cluster GCM}
\begin{figure}[htb]
\begin{center}
\includegraphics[width=6.0cm]{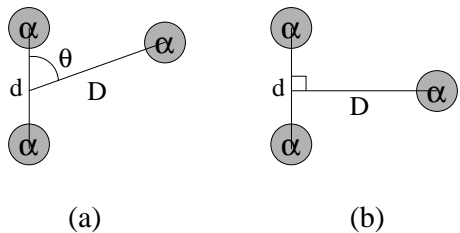} 	
\end{center}
  \caption{
Schematic figures of $3\alpha$ configurations for (a) the $3\alpha$GCM(full)
with the generator coordinates $d$, $D$, and $\theta$, and 
(b) the $3\alpha$GCM(radial)
with the generator coordinates $d$ and $D$, and the fixed $\theta=\pi/2$. 
\label{fig:3alpha}}
\end{figure}

To incorporate the large amplitude $\alpha$-cluster motions, I combine 
the $3\alpha$-cluster GCM($3\alpha$GCM) with the sAMD. 
In the $3\alpha$GCM, I superpose the $3\alpha$-cluster wave functions 
projected from the Brink-Bloch $3\alpha$-cluster wave functions \cite{Brink66}, 
\begin{equation}
\Phi_{3\alpha}=\frac{1}{\sqrt{A!}}{\cal A}\left\{\Phi_\alpha(\bvec{R}_1) \Phi_\alpha(\bvec{R}_2) \Phi_\alpha(\bvec{R}_3)    \right\},
\end{equation}
where $\Phi_\alpha(\bvec{R}_i) $ is the $\alpha$-cluster wave function written by the 
harmonic oscillator $(0s)^4$ configuration located at $\bvec{R}_i$ with the 
width $b=1/\sqrt{2\nu}$, and $\bvec{R}_i$ satisfies the relation $\bvec{R}_1+\bvec{R}_2+\bvec{R}_3=0$.
Note that $\Phi_{3\alpha}(\bvec{R}_1,\bvec{R}_2,\bvec{R}_3)$ can be expressed by the 
AMD wave function with the specific configuration.
To taken into account various $3\alpha$ configurations, the distances $d=|\bvec{R}_1-\bvec{R}_2|$ and 
$D=|\bvec{R}_3-(\bvec{R}_1+\bvec{R}_2)/2|$, and the angle $\theta$ between $\bvec{R}_3$ and 
$\bvec{R}_1-\bvec{R}_2$ are treated as generator coordinates
in the GCM calculation (see Fig.~\ref{fig:3alpha}(a)). Then, I combine the $3\alpha$GCM with the
sAMD and express the total wave function as
\begin{eqnarray}
\Psi^{\textrm{sAMD+}3\alpha\textrm{GCM}}_{^{12}\textrm{C}(J^\pi_k)}&&=
\sum_K c_0(J^\pi_k; K) 
P^{J\pi}_{MK}\Phi_{\rm AMD}({\bf Z}^0)\nonumber\\
&&+\sum_{i=1,\ldots,A}\sum_{\sigma}\sum_K 
c_1(J^\pi_k; i,\sigma,K) P^{J\pi}_{MK}\Phi_{\rm AMD}({\bf Z}_{\rm nonflip}^0(i,\sigma))\nonumber\\
&&+
\sum_{i=1,\ldots,A}\sum_{\sigma}\sum_K 
c_2(J^\pi_k; i,\sigma,K) P^{J\pi}_{MK}\Phi_{\rm AMD}({\bf Z}_{\rm flip}^0(i,\sigma))\nonumber  \\
&&+\sum_{d,D,\theta}\sum_K c_3(J^\pi_k; d,D,\theta,K) 
P^{J\pi}_{MK}\Phi_{3\alpha}(d,D,\theta)\label{eq:sAMD+aGCM},\label{eq:sAMD+GCM}
\end{eqnarray}
where the coefficients are determined by 
diagonalization of the norm and Hamiltonian matrices. The summation for three parameters 
$d$, $D$, $\theta$  corresponds to the $3\alpha$GCM with full $3\alpha$ configurations, which I call the
$3\alpha$GCM(full) in the present paper. To see the effect of the angular motion 
of $\alpha$ clusters, I also perform the $3\alpha$GCM calculation with the fixed angle $\theta=\pi/2$ considering only
the radial motion with the generator coordinates $d$ and $D$ (see Fig.~\ref{fig:3alpha}(b)), which I call 
$3\alpha$GCM(radial). 

\subsection{Isoscalar monopole and dipole transitions}

The ISM and ISD operators ${\cal M}(\textrm{IS0})$ and ${\cal M}(\textrm{IS1};\mu)$, 
which excite the compressional monopole and dipole modes, respectively, are
defined as
\begin{eqnarray}
{\cal M}(\textrm{IS0})&=&\sum_i  r^2_{i,\textrm{in}}, \\
{\cal M}(\textrm{IS1};\mu)&=&\sum_i  r^3_{i,\textrm{in}} Y^1_\mu(\hat{\bvec{r}}_{i,\textrm{in}}),
\end{eqnarray}
where $\bvec{r}_{i,\textrm{in}}$ is the $i$th nucleon coordinate with respect to the center of mass.
The ISM strength for $0^+_1\to 0^+_k$ and the ISD strength for  $0^+_1\to 1^-_k$ 
are given by the reduced matrix elements of these operators as
\begin{eqnarray}
B(\textrm{IS0};0^+_1\to 0^+_k)&=&\frac{1}{2J_\textrm{g.s.}+1}
|\langle 0^+_k||{\cal M}(\textrm{IS0})|| 0^+_1 \rangle |^2,\\
B(\textrm{IS1};0^+_1 \to 1^-_k)&=&\frac{1}{2J_\textrm{g.s.}+1}
|\langle 1^-_k||{\cal M}(\textrm{IS1})|| 0^+_1 \rangle |^2,
\end{eqnarray}
where  $J_\textrm{g.s.}$ is the ground state angular momentum, and in the present case
it is zero.
The energy-weighted sum (EWS) of the strengths is 
given as
\begin{eqnarray}
S(\textrm{IS0})\equiv \sum_{k\ge 2} E_{0^+_k} B(\textrm{IS0};0^+_1\to 0^+_k),\\
S(\textrm{IS1})\equiv \sum_{k\ge 1} E_{1^-_k} B(\textrm{IS1}; 0^+_1 \to 1^-_k).
\end{eqnarray}
If the interaction commutes with the ISM operator, the ISM energy weighted sum rule (EWSR) is given
as
\begin{eqnarray}
S(\textrm{IS0}) = \frac{2\hbar^2}{m}\langle 0^+_1|\sum_i r^2_{i,\textrm{in}}|0^+_1\rangle=
\frac{2\hbar^2}{m}A\langle r^2 \rangle_\textrm{g.s.}, \label{eq:eqsr}
\end{eqnarray}
where $\langle r^2 \rangle_\textrm{g.s.}$ is the mean-square radius of the ground state.

\section{Results}\label{sec:results}

\subsection{Effective nuclear interactions}
I use an effective nuclear interaction consisting of the central force of the MV1 force (case 3)\cite{TOHSAKI} and the
spin-orbit force of the G3RS force \cite{LS1,LS2}, and the Coulomb force.
The MV1 force consists of a two-range Gaussian two-body term and a zero-range three-body term.
The G3RS spin-orbit force is a two-range Gaussian force.
The Bartlett, Heisenberg, and Majorana parameters, $b=h=0$ and $m=0.62$,
in the MV1 force are adopted, and the strengths $u_{I}=-u_{II}\equiv u_{ls}=3000$ MeV 
of the G3RS spin-orbit force is used. These interaction parameters are the same
as those used in 
Refs.~\cite{KanadaEn'yo:1998rf,KanadaEn'yo:2006ze}, 
in which the AMD+VAP calculation describes 
properties of the ground and excited states in $E\lesssim 15$ MeV
of $^{12}$C.

\subsection{Parameter setting and the ground state of $^{12}$C}
The width parameter $\nu=0.19$ fm$^{-2}$ is chosen for the AMD wave function 
so as to minimize the ground state energy of $^{12}$C calculated 
by the AMD+VAP.

\begin{figure}[htb]
\begin{center}
\includegraphics[width=6.0cm]{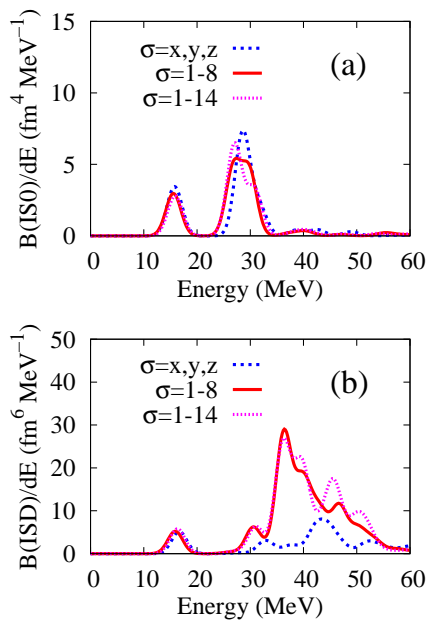} 	
\end{center}
  \caption{(color online) 
(a) ISM and (b) ISD strength distributions of $^{12}$C calculated by the sAMD 
with $\bvec{e}_{\sigma=x,y,z}$, $\bvec{e}_{\sigma=1,\ldots,8}$, and $\bvec{e}_{\sigma=1,\ldots,14}$.
The strengths are smeared by a Gaussian with the width $\gamma=2$ MeV.
\label{fig:c12-sp14}}
\end{figure}

In the sAMD, I try three sets of vectors, 
$\bvec{\Delta}_{\sigma=x,y,z}=\epsilon(1,0,0), \epsilon(0,1,0), \epsilon(0,0,1)$, 
$\bvec{\Delta}_{\sigma=1,\ldots,8}=\epsilon(\pm 1, \pm 1, \pm 1)$, 
$\bvec{\Delta}_{\sigma=1,\ldots,14}=\epsilon(\pm 1, \pm 1, \pm 1)$, 
$(\pm 1, 0, 0)$, $(0, \pm 1, 0)$, $(0, 0, \pm 1)$
for  the shift of Gaussian centroids, 
and check the convergence of the ISM and ISD strengths on the number of the directions $\sigma$.
Here, $\epsilon$ is chosen to be 0.1 fm which is 
small enough to give the $\epsilon$-independent result. 
The $x$, $y$, and $z$ axes are chosen to be the principle axes of the inertia in the intrinsic frame
and satisfy $\langle x^2\rangle \le \langle y^2\rangle\le \langle z^2\rangle$ 
and $\langle xy\rangle=\langle yz\rangle=\langle zx\rangle=0$ for the intrinsic wave function
$\Phi_{\rm AMD}({\bf Z}^0)$.
The ISM and ISD strengths calculated by the sAMD with three choices, 
$\bvec{e}_{\sigma=x,y,z}$, $\bvec{e}_{\sigma=1,\ldots,8}$, and $\bvec{e}_{\sigma=1,\ldots,14}$,
are shown in Fig.~\ref{fig:c12-sp14}, which shows the strength distributions 
smeared by a Gaussian with the width $\gamma=2$ MeV.
It is found that the set $\bvec{\Delta}_{\sigma=1,\ldots,8}$ is practically enough to get 
an approximately converged result for both the ISM and ISD strengths, whereas 
$\bvec{\Delta}_{\sigma=x,y,z}$ is enough only for the ISM strengths but not for the
ISD strengths. I adopt the set $\bvec{\Delta}_{\sigma=1,\ldots,8}$ in the present 
sAMD+$3\alpha$GCM calculation. 

For the generator coordinates in the $3\alpha$GCM, 
 $d=1.8+n_d$ fm ($n_d=0,\ldots,4$) and $D=2+n_D$ fm ($n_D=0,\ldots,5$) are adopted
in the $3\alpha$GCM(full) and $3\alpha$GCM(radial). In addition, $\theta=\pi n_\theta/8$ ($n_\theta=0,\ldots,4$)
are used in the $3\alpha$GCM(full).

\subsection{Ground state structure}

\begin{figure}[htb]
\begin{center}
\includegraphics[width=7.5cm]{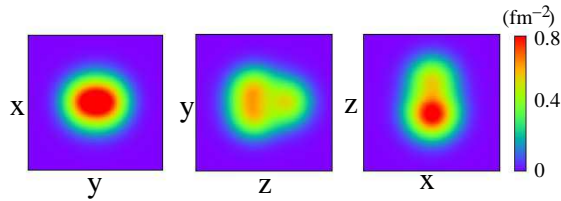} 	
\end{center}
  \caption{(color online) Density distribution of the intrinsic wave function of the $^{12}$C ground state
obtained by the AMD+VAP. The density is projected onto the $xy$(left), $yz$(middle), and $zx$(right) planes. 
\label{fig:c12-dense}}
\end{figure}

The wave function $\Phi_{\rm AMD}({\bf Z}^0)$ obtained by the AMD+VAP for the ground state contains 
the $\alpha$ cluster correlation as well as the $p_{3/2}$-closed configuration 
as discussed in Refs.~\cite{KanadaEn'yo:1998rf,KanadaEn'yo:2006ze}. 
As seen in the density distribution in Fig.~\ref{fig:c12-dense}, 
the intrinsic wave function of the ground state has a triaxial deformation with the cluster structure. 
In the sAMD and sAMD+$3\alpha$GCM(full)
calculations, the ground state wave function is expressed by the linear combination
of many configurations as given in Eqs.~\eqref{eq:sAMD} and \eqref{eq:sAMD+GCM}, 
however, it is still dominated by $P^{0+}_{00}\Phi_{\rm AMD}({\bf Z}^0)$ with 96\% (91\%)  in the sAMD(sAMD+$3\alpha$GCM(full)).
The calculated binding energy (B.E.) and root-mean-square matter radius ($r_m$) of $^{12}$C 
calculated  
by the AMD+VAP, sAMD, and sAMD+$3\alpha$GCM(full) are 
$\textrm{B.E.}=86.7$ MeV, 89.0 MeV, 89.6 MeV, and $r_m=2.41$ fm, 2.41 fm, and 2.46 fm, respectively, 
which reasonably agree to the experimental values
$\textrm{B.E.}=92.16$ MeV and $r_p=2.33$ fm (the point-proton radius) reduced from the 
charge radius \cite{Angeli2013}.

\subsection{ISM strengths}
\begin{figure}[htb]
\begin{center}
\includegraphics[width=6.0cm]{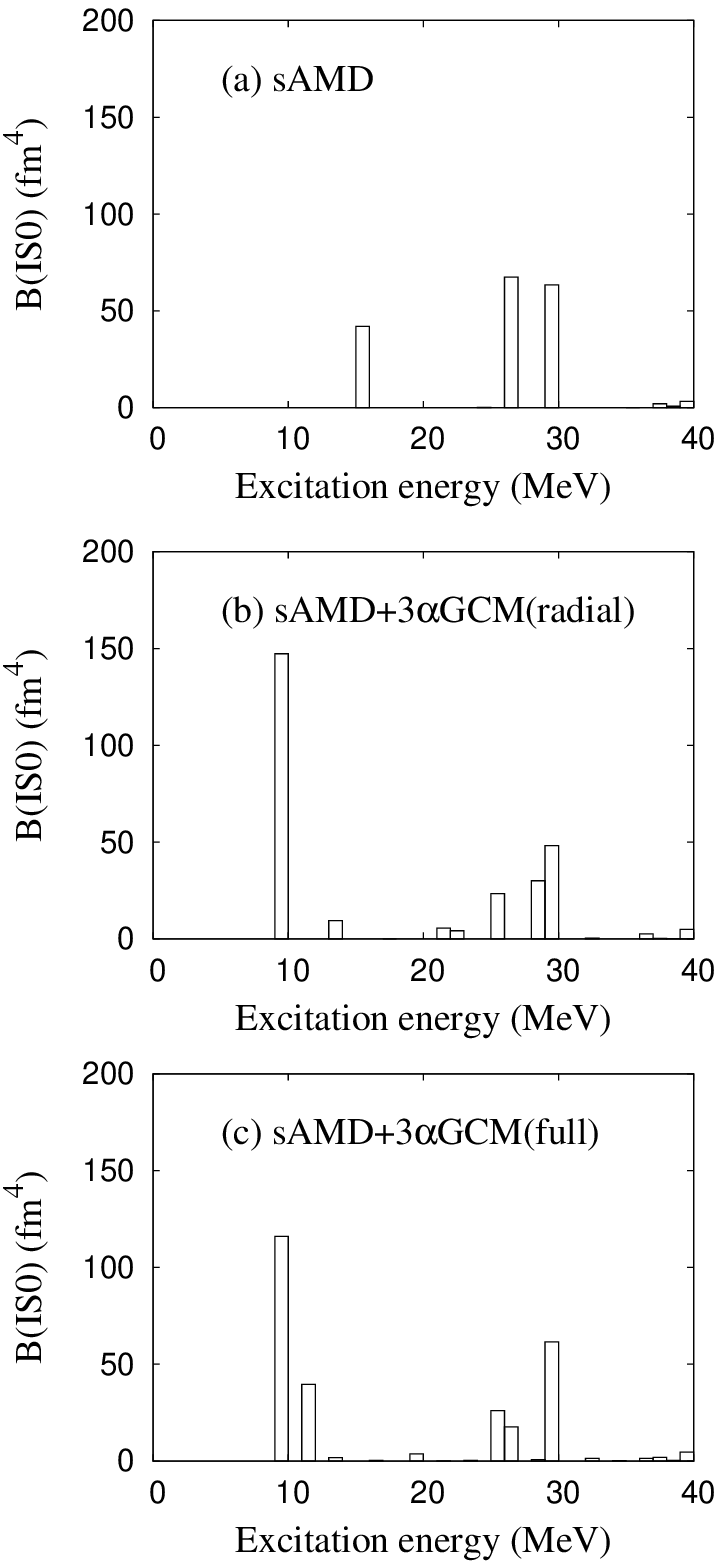} 	
\end{center}
  \caption{The ISM strengths calculated by the sAMD, sAMD+$3\alpha$GCM(radial), 
and sAMD+$3\alpha$GCM(full). The sum of the strengths $B(\textrm{IS0})$ 
in each energy bin is shown. 
\label{fig:c12-is0}}
\end{figure}

The ISM strengths for $0^+_1\to 0^+_k$ ($k\ge 2$) 
obtained by the sAMD, sAMD+$3\alpha$GCM(radial), 
and sAMD+$3\alpha$GCM(full)  are shown in Fig.~\ref{fig:c12-is0}.
In the sAMD, sAMD+$3\alpha$GCM(radial), and 
sAMD+$3\alpha$GCM(full) results, 
the total energy weighted sum (TEWS) of the ISM strengths
corresponds to 97\%, 100\%, and 100\% of the
EWSR in Eq.~\eqref{eq:eqsr}, respectively.
In the three calculations, the remarkable ISM strengths exist in $E=25-30$ MeV corresponding to
the ISGMR.
In the sAMD result (Fig.~\ref{fig:c12-is0}(a)), 
an low-energy ISM resonance is found at $E\sim 15$ MeV 
and regarded as a signal of the cluster excitation. 
 As seen in the sAMD+$3\alpha$GCM(radial) result (Fig.~\ref{fig:c12-is0}(b)), 
this low-energy mode comes down below $E=10$ MeV, and its ISM strength is remarkably 
enhanced by the large amplitude radial motion of $\alpha$ clusters. It should be noted that 
a part of the ISGMR strength feeds the low-energy strength. 
Furthermore, in the sAMD+$3\alpha$GCM(full) calculation with 
the radian and angular motions of $\alpha$ clusters,
the low-energy mode splits into two states around $E=10$ MeV (see Fig.~\ref{fig:c12-is0}(c)).

The lower state is assigned to the experimentally known 
$0^+_2$ state at 7.65 MeV.  The calculated $B(\textrm{IS0})$ agrees well to the experimental value 
$B(\textrm{IS0})=120\pm 4$ fm$^4$
evaluated from the E0 strength measured by electron scatterings \cite{Chernykh:2010zu}
as $B(\textrm{IS0})=4B(\textrm{E0})$ assuming the isospin symmetry. 
The higher state corresponds to the $0^+_3$ state predicted by the AMD and FMD calculations 
\cite{KanadaEn'yo:1998rf,Neff-c12,KanadaEn'yo:2006ze,Chernykh:2007zz} as well as the $3\alpha$-cluster calculations \cite{uegaki1,Arai:2006bt}.
The sum of the ISM strengths of the two $0^+$ states ($0^+_2$ and $0^+_3$) in the sAMD+$3\alpha$GCM(full) is almost the same as the ISM strength of the low-energy mode obtained by the sAMD+$3\alpha$GCM(radial). Namely, the low-energy mode with the 
remarkable ISM strength arises from the the radial excitation of $\alpha$ clusters, and its strength is dominantly 
carried by the $0^+_2$ and partly shared by the $0^+_3$ state.
In the comparison of the ISM strengths between the sAMD, sAMD+$3\alpha$GCM(radial), and sAMD+$3\alpha$GCM(full) results, it is found 
that the enhancement and splitting of the low-energy ISM strengths 
occur from the radial and angular motions of $\alpha$ clusters, respectively.  

\subsection{ISD strengths}
\begin{figure}[htb]
\begin{center}
\includegraphics[width=6.0cm]{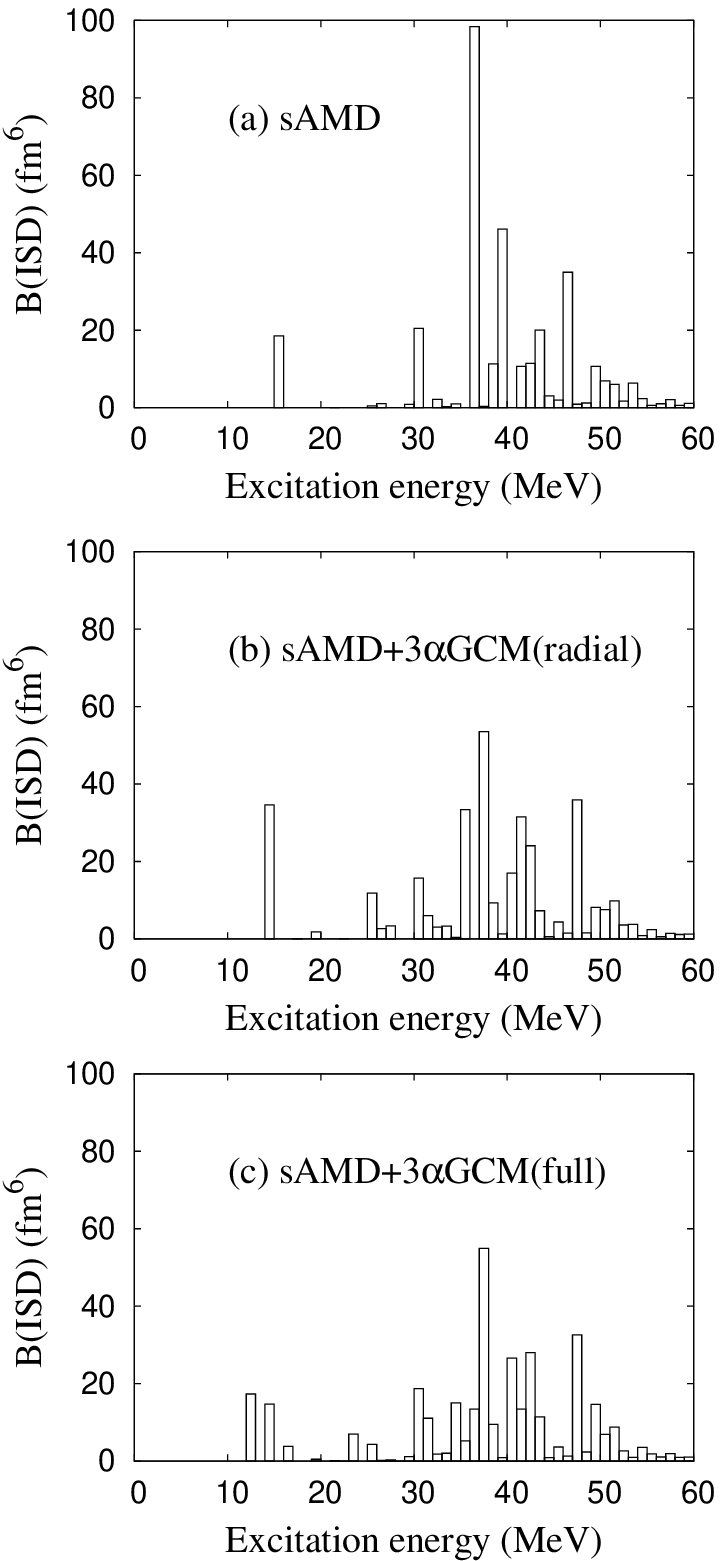} 	
\end{center}
  \caption{The ISD strengths calculated by the sAMD, sAMD+$3\alpha$GCM(radial), 
and sAMD+$3\alpha$GCM(full). The sum of the strengths $B(\textrm{IS1})$ in  each energy bin is shown. 
\label{fig:c12-isd}}
\end{figure}

The ISD strengths for $0^+_1\to 1^-_k$ ($k\ge 1$) 
calculated by the sAMD, sAMD+$3\alpha$GCM(radial), 
and sAMD+$3\alpha$GCM(full) are shown in Fig.~\ref{fig:c12-isd}.
Interestingly, also in the ISD strengths, 
the enhancement and splitting of the low-energy strengths 
occur from the radial and angular motions of $\alpha$ clusters, respectively.
The ISGDR is found in the energy $E=30-55$ MeV region  higher than the ISGMR energy. 
These high-energy strengths for the ISGDR are not so affected by the large amplitude cluster motions. 
Below the ISGDR, the sAMD result shows an low-energy ISD resonance at $E\sim 15$ MeV regarded as 
a signal of the cluster mode.
The low-energy ISD strength of the cluster mode is enhanced about twice 
by the radial motion of $\alpha$ clusters in the sAMD+$3\alpha$GCM(radial) result, 
and it splits into at least two states in $E=10-15$ MeV because of the angular motion of $\alpha$ clusters
in the sAMD+$3\alpha$GCM(full) result. Consequently, 
the low-energy ISD strengths of the cluster mode are shared by the two states almost equally. 
The lower state can be assigned to the experimentally known $1^-_1$ state at 10.84 MeV. 
For the higher $1^-$ state, there is no corresponding state confirmed experimentally, 
however, the significant ISD strengths around $E=15$ MeV have been observed by $(\alpha,\alpha')$ 
scatterings and could be a signal of the higher $1^-$ state obtained in the present calculation. More details are discussed later. 

\subsection{Energy weighted strength distributions}

\begin{figure}[htb]
\begin{center}
\includegraphics[width=6.0cm]{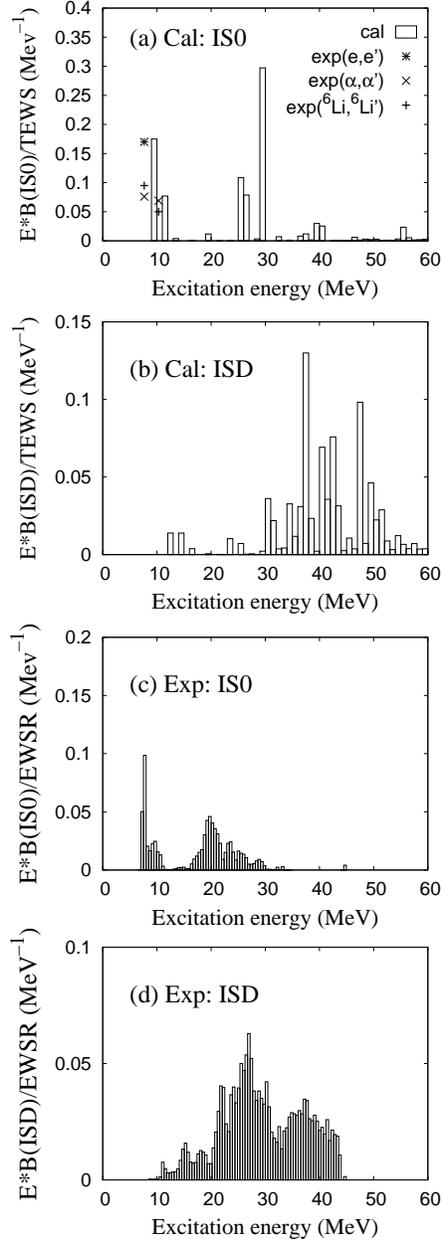} 	
\end{center}
  \caption{The energy weighted ISM and ISD strength distributions 
calculated by the sAMD+$3\alpha$GCM(full) and those 
measured by $(\alpha,\alpha')$ scatterings. 
For the calculated result, the ratio of the strengths in each energy bin 
to the TEWS is shown. 
The experimental data are those to the EWSR \cite{John:2003ke}.
The panel (a) also shows the experimental values of the energy weighted 
ISM strengths in the unit of EWSR: 
the data for the $0^+_2$ and $0^+_3$ states measured by $(\alpha,\alpha')$\cite{John:2003ke}
and ($^6$Li,$^6$Li$'$) \cite{Eyrich:1987zz} scatterings,
and that evaluated by the E0 strength measured by $(e,e')$ scatterings \cite{Chernykh:2010zu}
assuming the mirror symmetry.
\label{fig:c12-ewsr}}
\end{figure}

The energy weighted ISM and ISD strength distributions 
calculated by the sAMD+$3\alpha$GCM(full) are compared with those 
measured by $(\alpha,\alpha')$ scatterings \cite{John:2003ke} in Fig.~\ref{fig:c12-ewsr}. 
The calculated high-energy strengths for the 
ISGMR and ISGDR are found around $E=30$ MeV and $E=40$ MeV, respectively.  
The low-energy ISM and ISD strengths for the cluster modes
exist in the energy regions much below the giant resonances. 
The percentages of the strengths in the low-energy and high-energy regions calculated by the 
sAMD+$3\alpha$GCM(full) are shown in Table \ref{tab:EWS}, compared with those of 
the sAMD and sAMD+$3\alpha$GCM(radial) calculations.

The low-energy ISM strengths for the $0^+$ states
around $E=10$ MeV is only 11\% of the EWSR in the sAMD, 
however, they are largely enhanced by the large amplitude cluster motions and 
exhaust 26\% of the EWSR in the sAMD+$3\alpha$GCM(radial) 
and sAMD+$3\alpha$GCM(full). The fact that the sum of the 
low-energy and high-energy strengths is almost constant ($75-77$ \% of the EWSR) 
in three calculations indicates that the increment of the low-energy ISM strengths 
come from the ISGMR strength, which originally concentrates in the high-energy region in 
the sAMD. In other words, in the sAMD calculation without the large 
amplitude cluster modes, a part of the strengths of the cluster modes is involved by 
the ISGMR. 

The feature of the experimental ISM strength distributions (Fig.~\ref{fig:c12-ewsr} (c)) which 
consists of the high-energy strengths of the ISGMR and the significant 
low-energy strengths are qualitatively described by the calculation (Fig.~\ref{fig:c12-ewsr}(a)),
however, the quantitative description of the high-energy part is 
not satisfactory in the present result.
The present calculation overestimates 
the centroid energy (21.9$\pm$0.3 MeV) and the strengths (27$\pm$5\% of the EWSR)
of the observed ISGMR, and 
also fails to describe the width of the ISGMR. 
The present model space of the sAMD is restricted only in 
the single-particle excitations written by shifted Gaussians and insufficient 
to describe the spreading width of the ISGMR.

For the low-energy ISM strengths,
the calculated strength for the $0^+_2$ state exhausts 18\% of the EWSR,
which well agrees to the experimental values for the $0^+_2$(7.65 MeV), 
15\% in Ref.~\cite{strehl70} and 
17\% in Ref.~\cite{Chernykh:2010zu}, measured 
by $(e,e')$ scatterings (see Fig.~\ref{fig:c12-ewsr}(a) and Table \ref{tab:12C0}). 
The data evaluated by 
$(\alpha,\alpha')$\cite{John:2003ke} and ($^6$Li,$^6$Li$'$) \cite{Eyrich:1987zz}scatterings 
is 7.6$\pm$0.9\% and 9.5\%, which is about a half of the values
measured $(e,e')$ scatterings. This inconsistency of the ISM strength for the $0^+_2$ state
might come from the reaction model dependence in the DWBA analysis of
nuclear scatterings.
For the $0^+_3$ state,
the calculated strength exhausts 8\% of the EWSR,
which reasonably agrees to the experimental values, 6.9$\pm$0.9\% and 5$\pm$1\%, 
for the $0^+_3$(10.3 MeV) 
measured by the $(\alpha,\alpha')$
and ($^6$Li,$^6$Li$'$) scatterings, respectively.
The observed ISM strengths around $E=10$ MeV were considered as a board $0^+_3$ state 
at $E=10.3$ MeV with 
a width of $\Gamma\sim 3$ MeV \cite{John:2003ke}. However, the recent experiment reported 
an indication that it contains two $0^+$ states at $E=9.04 \pm 0.09$ MeV with $\Gamma=1.45\pm 0.18$ 
and $E=10.56\pm 0.06$ MeV with $\Gamma=1.42\pm 0.08$ MeV \cite{Itoh:2011zz}.
It is likely that the ISM strengths for the $0^+_3$(10.3 MeV) reported in 
Refs.~\cite{John:2003ke,Eyrich:1987zz} may contain the strengths for the two $0^+$ states
around $E=10$ MeV above the $0^+_2$.
Theoretically, the 3$\alpha$ orthogonal condition model ($3\alpha$OCM) calculations 
predicted  two $0^+$ states 
above the $0^+_2$ state \cite{Kurokawa:2005ax,ohtsubo13}, 
whereas the AMD, FMD, and microscopic 3$\alpha$-cluster calculations 
predicted only one $0^+$ state in this energy region
\cite{uegaki1,KanadaEn'yo:1998rf,Neff-c12,KanadaEn'yo:2006ze,Arai:2006bt,Chernykh:2007zz}. 
According to the 3$\alpha$OCM calculations, 
one of the two $0^+$ states is a very broad state.
The present framework is a bound state approximation, and therefore, it is not suitable 
to describe such the state strongly coupled by $3\alpha$ continuum. 
If the missing $0^+$ state comes to this energy region and mix with the 
$0^+_3$ state, it could share a portion of the ISM strength for the $0^+_3$ 
obtained in the present calculation.
It means that the calculated ISM strength for the $0^+_3$ exhausting 8\% of the EWSR
may correspond to the sum of the ISM strengths for possible $0^+$ states 
in this energy region consistently to the experimental 
strengths evaluated by assuming a single $0^+$ state at $E\sim 10$ MeV in Refs.~\cite{John:2003ke,Eyrich:1987zz}.

Let me look into the ISD strength distributions. 
As seen in Fig.~\ref{fig:c12-ewsr}(b), 
the high-energy ISD strengths corresponding to the 
ISGDR show a broad distribution mainly in $E=30-55$ MeV.
The structure of the broad ISGDR is qualitatively consistent with the observed
ISD strengths shown in Fig.~\ref{fig:c12-ewsr}(d), however, the present calculation 
overestimates the observed peak energy of the ISGDR by about 10 MeV.
The calculated ISD strengths in $E<55$ MeV is 86\% of the TEWS in the 
sAMD+$3\alpha$GCM(full), 
which is consistent with the observed ones in $E<45$ MeV exhausting 
78$\pm$9\% of the EWSR \cite{John:2003ke}.

The low-energy ISD strengths between  $E=10-20$ MeV exhaust 3.4\% of the TEWS 
in the sAMD+$3\alpha$GCM(full). The calculated low-energy ISD strengths in $^{12}$C 
is comparable to the observed low-energy ISD strengths 
(4\% of the EWSR) in $^{16}$O \cite{Harakeh:1981zz}. 
Compared with the sAMD result, the strength is  
enhanced by a factor 1.5 in the sAMD+$3\alpha$GCM(full) 
because of the cluster components in both the 
ground state and excited $1^-$ states.
A couple of $1^-$ states contribute to the low-energy ISD strengths 
as seen in Fig.~\ref{fig:c12-ewsr}(b) for the sAMD+$3\alpha$GCM(full).
The excitation energies of $E=12.6$ MeV and $E=14.8$ MeV of 
the $1^-_1$ and $1^-_2$ states obtained by the sAMD+$3\alpha$GCM(full)
are consistent with those suggested by the $3\alpha$-cluster calculation in Ref.~\cite{uegaki1}.
The $1^-_1$ state is assigned to the experimentally known $1^-$ state at 10.84 MeV. 
For the $1^-_2$ state, there is no experimentally confirmed state.
The calculated ISD strength for the $1^-_2$ state is 1.5\% of the TEWS. 
The significant ISD strength suggests that the ISD excitations 
can be a good probe to experimentally observe the $1^-_2$ state. 
In the observed ISD strengths shown in Fig.~\ref{fig:c12-ewsr}(d),  one can see a
bump at $E\sim 15$ MeV, which can be a candidate of the $1^-_2$ state obtained in the
present calculation. Unfortunately, it is difficult to extract the ISD strength for this state 
from the data because of the back ground ambiguity. 
The existence of the significant ISD strengths at $E\sim 15$ MeV is also supported by 
the measurement on $(\alpha,\alpha')$ scattering at 386 MeV by Itoh {\it et al.} \cite{Itoh15}. 

\begin{table}[htb]
\caption{The energy weighted ISM and ISD strengths
in the low-energy and high-energy regions 
calculated by the sAMD, sAMD+$3\alpha$GCM(radial), and 
sAMD+$3\alpha$GCM(full). 
The percentages of the ISM strengths in $E<17$ MeV and 
$17<E<45$ MeV to the EWSR and 
those of the ISD strengths in $E<20$ MeV and $20<E<55$ MeV 
to the TEWS are listed. The ISM TEWS  (fm$^4$MeV), ISM EWSR (fm$^4$MeV), 
and ISD TEWS (fm$^6$MeV) are also shown.
\label{tab:EWS} 
 }
\begin{center}
\begin{tabular}{ccccc}
\hline
& sAMD & \multicolumn{2}{c}{sAMD+$3\alpha$GCM} \\
&   &  (radial)  & (full) \\
\hline
IS0 &  & & \\
TEWS& $5.6\times 10^3$ & $6.0\times 10^3$  & $6.0\times 10^3$ \\
EWSR&  $5.8\times 10^3$ & $6.0\times 10^3$  & $6.0\times 10^3$ \\
$E<17$  & 11\% &  26\% & 26\% \\ 
$17<E<45$ & 0.64\% &	0.51\% &	50\% \\
IS1 & & & \\
TEWS & $1.41\times 10^4$ &$1.45\times 10^4$ & $1.46\times 10^4$\\
$E<20$ & 2.1\%& 	0.03.6\% &	3.4\%\\ 
$20<E<55$ & 87\%& 	83\%& 	83\% \\
\hline
\end{tabular}
\end{center}
\end{table}

\begin{table}[htb]
\caption{The energy-weighted ISM strengths for the $0^+_2$ and $0^+_3$ states.
The experimental data are those measured by $(\alpha,\alpha')$\cite{John:2003ke}
and ($^6$Li,$^6$Li$'$) \cite{Eyrich:1987zz} scatterings, and those evaluated by the
E0 strengths measure by $(e,e')$ \cite{strehl70,Chernykh:2010zu} assuming the mirror
symmetry.  The theoretical values are calculated by the sAMD+$3\alpha$GCM(full). 
The percentages to the EWSR are shown.
\label{tab:12C0} 
 }
\begin{center}
\begin{tabular}{cccccc}
\hline
& cal. & $(e,e')$ \cite{strehl70} & $(e,e')$\cite{Chernykh:2010zu} & $(\alpha,\alpha')$ \cite{John:2003ke}
& ($^6$Li,$^6$Li$'$) \cite{Eyrich:1987zz}\\
\hline
$0^+_2$ & 18 & 15 & 17 & 7.6$\pm$0.9 &	9.5 \\
$0^+_3$ &  8  &     &      &   6.9$\pm$0.9 & 5$\pm$1  \\
\hline
\end{tabular}
\end{center}
\end{table}

\section{Summary} \label{sec:summary}
I investigated the ISM and ISD excitations in $^{12}$C
with the sAMD+$3\alpha$GCM calculation. The small amplitude vibration modes are 
described by the coherent 1p-1h excitations expressed by small shift of single-nucleon 
Gaussian wave functions, whereas the large amplitude cluster modes are incorporated 
by superposing $3\alpha$-cluster wave functions in the GCM.
The coupling of the excitations
in the intrinsic frame with the rotation and parity transformation is taken into account 
microscopically by the angular-momentum and parity projections.
The present calculation describes the ISM and ISD excitations 
in a wide energy region covering the cluster excitations in the low-energy region 
and the giant resonances in the high-energy region. 

In the calculated ISM strengths, the significant strengths corresponding 
to the cluster excitations are obtained in the low-energy region below the ISGMR.
The low-energy ISM strength of the cluster mode is remarkably 
enhanced by the radial motion of $\alpha$ clusters, and it splits into
two states by the angular motion of $\alpha$ clusters. The low-energy ISM strengths
exhaust 26\% of the EWSR, which is consistent with the experimental data
for the $0^+$ states at 7.65 MeV and 10.3 MeV
measured by $(e,e')$, $(\alpha,\alpha')$, and ($^6$Li,$^6$Li$'$) 
scatterings.
The feature of the experimental ISM strength distributions which 
consists of the high-energy strengths of the ISGMR and the significant 
low-energy strengths are qualitatively described by the calculation,
however, the quantitative description of the high-energy part is 
not satisfactory in the present result. 
The present calculation overestimates 
the centroid energy (21.9$\pm$0.3 MeV) and the strengths (27$\pm$5\% of the EWSR)
of the ISGMR observed by $(\alpha,\alpha')$ scatterings \cite{John:2003ke}, and 
also fails to describe the experimental width of the ISGMR. 
The present model space is restricted only in
the single-particle excitations written by shifted Gaussians and insufficient 
to describe the spreading width of the ISGMR.

Also in the calculated ISD strengths, 
the low-energy strength of the cluster mode is 
enhanced by the radial motion of $\alpha$ clusters, and it splits into
a couple of states because of the angular motion of $\alpha$ clusters.
Consequently, two  $1^-$ states (the $1^-_1$ and $1^-_2$) 
having the significant ISD strengths are obtained in $E=10-15$ MeV. 
The lower state is assigned to the experimentally known $1^-$ state at 10.84 MeV. 
For the $1^-_2$ state, there is no experimentally confirmed state.
The present calculation indicates that the ISD excitations 
can be a good probe to experimentally observe the $1^-_2$ state. 
In the experimental ISD strengths measured by $(\alpha,\alpha')$ scatterings \cite{John:2003ke,Itoh15},
the bump observed at $E\sim 15$ MeV can
be a candidate of the $1^-_2$ state obtained in the
present calculation.

\section*{Acknowledgments} 
The author would like to thank Dr.~Itoh and Dr. Kimura for fruitful discussions.
The computational calculations of this work were performed by using the
supercomputer in the Yukawa Institute for theoretical physics, Kyoto University. This work was supported by 
JSPS KAKENHI Grant Number 26400270.

\end{document}